Title: **A 2D autocorrelation-based frequency estimator reflecting spatial tissue distribution to improve Ultrasound H-scan tissue characterization**

Authors:

Jihye Baek[1], Thurston Brevett[1], Dongwoon Hyun[1], Ahmed El Kaffas[2], Aya Kamaya[1], Kevin J. Parker[3], and Jeremy J. Dahl[1]

Affiliations:

[1]Department of Radiology, Stanford University School of Medicine, Stanford, CA, USA

[2]Department of Radiology, University of California San Diego, La Jolla, CA, USA

[3]Department of Electrical and Computer Engineering, University of Rochester, Rochester, NY, USA

Corresponding author: Jeremy J. Dahl
jjdahl@stanford.edu

# ABSTRACT

**Abstract**— H-scan is a promising quantitative ultrasound technique that estimates the frequency content of backscattered signals and maps the estimated frequencies onto a red/blue color scale to reflect underlying tissue properties. Although it relies on matched filters tuned to different frequencies, the broad spectral bandwidth of ultrasound produces noisy, granular displays. Here, we introduce an adaptive frequency estimator designed to suppress the noise within homogeneous regions while preserving sharpness across tissue boundaries.

The method combines 2D autocorrelation with a matched filter. In the first stage, a matched-filter-based estimation yields an a priori map of the spatial distribution of frequencies. The local heterogeneity of these estimates then defines a 2D weighting function that guides a second estimation stage. Drawing on the concept of Loupas's blood velocity estimator, we apply autocorrelation over a 2D spatial kernel to recover the axial frequency components, employing a weighted summation that accounts for the spatial frequency distribution within the kernel. We benchmarked the proposed estimator against conventional approaches, including the short-time Fourier transform, the H-scan matched filter, and standard autocorrelation, using both Field II simulations and *in vivo* data from human subjects with hepatic steatosis.

In simulation, our adaptive estimator reduced the noisy texture in homogeneous regions while retaining clear boundary delineation, whereas the other estimators could achieve only one of these objectives. Applied to the *in vivo* human liver, the estimator improved H-scan image quality by lowering noise and enhancing the discrimination of steatotic liver from adjacent gallbladder and skin layers.

# 1. INTRODUCTION

Ultrasound tissue characterization techniques enhance disease diagnosis by providing clinicians with more comprehensive information beyond traditional B-mode ultrasound imaging. While B-mode imaging is valuable for disease screening due to the advantages of ultrasound—such as easy accessibility, low cost, real-time imaging, and portability—diagnosis based solely on B-mode imaging has limitations. Traditional approaches to quantitative ultrasound (QUS) have been developed that play a crucial role in disease diagnosis [1]. QUS can extract more specific features from tissues that surpass the qualitative insights provided by B-mode imaging, thereby offering additional diagnostic evidence. Initial efforts to quantify B-mode imaging included estimating B-mode brightness and the hepatic renal index (HRI) to aid in the diagnosis of steatosis. More sophisticated methods for obtaining quantitative measurements have also been explored. For instance, ultrasound radiofrequency (RF) data can be processed to measure spectral features such as midband fit and spectral slope, while speckle statistics can be analyzed using Rayleigh, homodyned-K, Nakagami, and Burr distributions to obtain distribution-related parameters. Additionally, ultrasound physics can be employed to estimate attenuation and backscatter coefficients. Furthermore, shear wave elastography, by manipulating the ultrasound transmission sequence from B-mode imaging, enables the evaluation of tissue stiffness through parameters such as shear wave speed, attenuation, and dispersion. These shear wave parameters provide critical evidence for screening liver fibrosis via ultrasound. Recently, local sound speed and H-scan have been proposed as novel tissue characterization approaches, showing potential for use in diagnosing hepatic steatosis.

H-scan has emerged as a promising novel approach in tissue characterization, allowing for improved identification of tissues compared to other quantitative ultrasound (QUS) measures, as evidenced by findings from previous studies involving various diseases in animal models and human subjects. In a simple animal model of Metabolic Dysfunction-Associated Steatotic Liver Disease (MASLD), H-scan demonstrated its ability to differentiate among the stages of steatosis: normal, early, and late [2]. It facilitated the differentiation of early steatosis from normal tissue, thereby enabling the monitoring of steatosis progression in the proposed imaging mode [3]. Moreover, H-scan applied to a combination of multiple liver diseases, including steatosis, fibrosis, and inflammation, demonstrated its capability to differentiate among these conditions [4, 5]. The progression of pancreatic cancer metastasis in the liver was also monitored by H-scan, showing better tracking than bioluminescent imaging and shear wave elastography [6]. Beyond its application in liver diseases, H-scan has been investigated in other clinical conditions. For instance, kidney fibrosis was successfully differentiated from normal tissue [7]. In a mouse model of melanoma, H-scan enabled the discrimination of tumor microenvironmental heterogeneity [8, 9]. Furthermore, H-scan has been applied to human subjects *in vivo*. It accurately monitored the progression of hepatic steatosis, and the H-scan parameters, such as color level and attenuation, exhibited a higher correlation with the current standard of MRI-PDFF than other QUS parameters [10]. For more complicated liver diseases, such as metabolic dysfunction-associated steatohepatitis (MASH), H-scan contributed to monitoring MASH progression [11]. Additionally, in human breast lesions, H-scan outperformed other QUS measures in effectively distinguishing between benign and malignant conditions [12].

Despite the promising potential of H-scan in tissue characterization, several limitations within this framework remain to be addressed. The original H-scan technique utilized matched filtered RF data with two filters, one for low frequency and the other for high frequency, and assigned the outputs to the RGB channels to generate color-coded images [13]. Recently, an optimized version of H-scan has been introduced, utilizing 256 matched filters to achieve a more diverse spectral composition, which allows for better representation of tissue signatures. Instead of employing RGB channel assignment, it utilized the index of the best-matched filter for color coding in the imaging process[14]. The index reflects the peak frequency of each matched filter, so this color coding captures the measured frequency components. Due to the broadband frequency spectrum of ultrasound echoes, visualizing H-scan as a local frequency map produces a noisy texture pattern. Thus, H-scan employs an almost binarized red/blue colormap to emphasize gross shifts resulting from pathological changes, rather than focusing on subtle local frequency variations. Nevertheless, enhancing the accuracy of H-scan measurements will advance tissue characterization.

To enhance tissue characterization, we focus on improving frequency estimation within the H-scan concept, specifically through the color-coding of frequency components. We explored other methodologies for local frequency estimation, including the Short-time Fourier transform (STFT), matched filter analysis (e.g., H-scan[14]), and axial autocorrelation, adapted from Loupas's 2D autocorrelator for blood flow [15].

For the STFT and 2D autocorrelator, it is essential to select an appropriate window size for the estimation kernel, whereas a window is not required for the matched filter. The noisy texture pattern generated by these estimators can be mitigated through low-pass filtering or by using a larger window in the STFT/autocorrelator; however, this comes with a trade-off between the variance of the estimator and its resolution. In this study, we propose an adaptive frequency estimator that combines the high resolution of the matched filter with the lower variance characteristic of the 2D autocorrelation method. We assessed the estimator's performance via Field

II simulation and an in vivo study, comparing its effectiveness against the STFT, H-scan matched filter, and 2D autocorrelator.

## 2. THEORY

### 2.1. H-scan for Tissue Characterization

H-scan is a promising framework for tissue characterization that utilizes matched filter analysis [14]. To quantify tissue signatures indicative of pathological conditions, the outputs from the matched filters are analyzed and color-coded, resulting in color images that are dependent on tissue conditions. This provides clinicians with valuable visual insights that aid in disease diagnosis. In particular, H-scan reveals color shifts from normal tissue states. The colors in H-scan are normalized to black for normal tissues; as pathological changes occur due to disease progression, the colors shift toward red or blue. Ideally, different diseases exhibit distinct directional shifts toward red or blue, allowing for the differentiation of tissues based on their specific conditions.

The H-scan utilizes ultrasound RF data as input, which are convolved in the axial direction using matched filters having different peak frequencies ranging from low to high at equal intervals to detect different scatterer sizes. Ultrasound propagation leads to frequency shifts of RF data due to attenuation or tissue conditions. To eliminate the frequency downshifts induced by attenuation with depth, we preprocessed the RF data for attenuation correction; more details for attenuation correction can be found in [14]. The attenuation-corrected RF data were then used for matched filtering. The H-scan generally employs 256 matched filters to provide color imaging with a resolution of 256 color scales or levels. Each matched filter is a Gaussian function, which models the ultrasound broadband spectrum, with a specific peak frequency. The matched filters have peak frequencies range from $f_0 - f_\Delta$ to $f_0 + f_\Delta$ MHz at equal interval where $f_0$ and $f_\Delta$ can be determined based on ultrasound transmission frequency and bandwidth of the spectrum, respectively. Each pixel or sample of RF data has 256 convolution values, with the maximum among the 256 indicating the dominant frequency component, which corresponds to a specific scatterer size. This is color-coded using the H-scan red-blue colormap. In H-scan imaging, greater blue shifts indicate relatively smaller scatterers, while greater red shifts correspond to larger scatterers.

### 2. 2. Two-Dimensional Autocorrelation for Blood Flow Velocity

Color Doppler imaging displays color coding to blood flow velocity based on frequency measures in the ensemble (time-series) direction using autocorrelation $\hat{\gamma}$ [16], defined as follows:

$$\hat{\gamma}(n') = \sum_{n=0}^{N-n'-1} \hat{r}(n)\hat{r}^*(n+n') \quad \text{(eq 1)}$$

Where $\hat{r}(n)$ is a discrete analytic signal and $n$ represents the ensemble direction. The mean Doppler frequency $\langle F \rangle_{1D}$ is given by:

$$\langle F \rangle_{1D} \cong \frac{1}{2\pi} \tan^{-1}\left\{\frac{Im[\hat{\gamma}(1)]}{Re[\hat{\gamma}(1)]}\right\}. \quad \text{(eq 2)}$$

More precise frequency measurement for blood flow velocity can be achieved a two-dimensional frequency estimation [15], which incorporates the axial directional frequency component along with ensemble direction utilizing 2D autocorrelation, given by:

$$\hat{\gamma}(m', n') = \sum_{m=0}^{M-m'-1} \sum_{n=0}^{N-n'-1} \hat{r}(m,n)\hat{r}^*(m+m', n+n') \quad \text{(eq 3)}$$

where $\hat{r}(m,n)$ is a discrete analytic signal:

$$\hat{r}(m,n) = I(m,n) + jQ(m,n). \quad \text{(eq 4)}$$

$n'$ and $n$ represent the ensemble direction with pulse repetition period, and $m'$ and $m$ represent the axial direction with sampling interval. Let define $H(m,n)$ as:

$$H(m,n)|_{m',n'} \equiv \hat{r}(m,n)\hat{r}^*(m+m',n+n') \quad \text{(eq 5)}$$

at given $m'$ and $n'$, and then $\hat{\gamma}(m',n')$ can be written by:

$$\hat{\gamma}(m',n') = \sum_{m=0}^{M-m'-1} \sum_{n=0}^{N-n'-1} H(m,n)|_{m',n'}$$

$$= \sum_{m=0}^{M-m'-1} \sum_{n=0}^{N-n'-1} \{Re[H(m,n)|_{m',n'}] + Im[H(m,n)|_{m',n'}]\}. \quad \text{(eq 6)}$$

The autocorrelation is intuitively interpreted by vector sum of $H(m,n)|_{m',n'}$ in real and imaginary axes as illustrated in **Figure 1 (a)**, showing that $H(m,n)|_{m',n'}$ vectors are equally summed within the rectangular region of interest (ROI) $(\{(m,n)|m = 0,1,\dots,M-m'-1 \ and \ n = 0,1,\dots,N-n'-1\})$ to calculate $\hat{\gamma}(m',n')$ without weighting specific subset of $\{H(m,n)|_{m',n'}\}$.

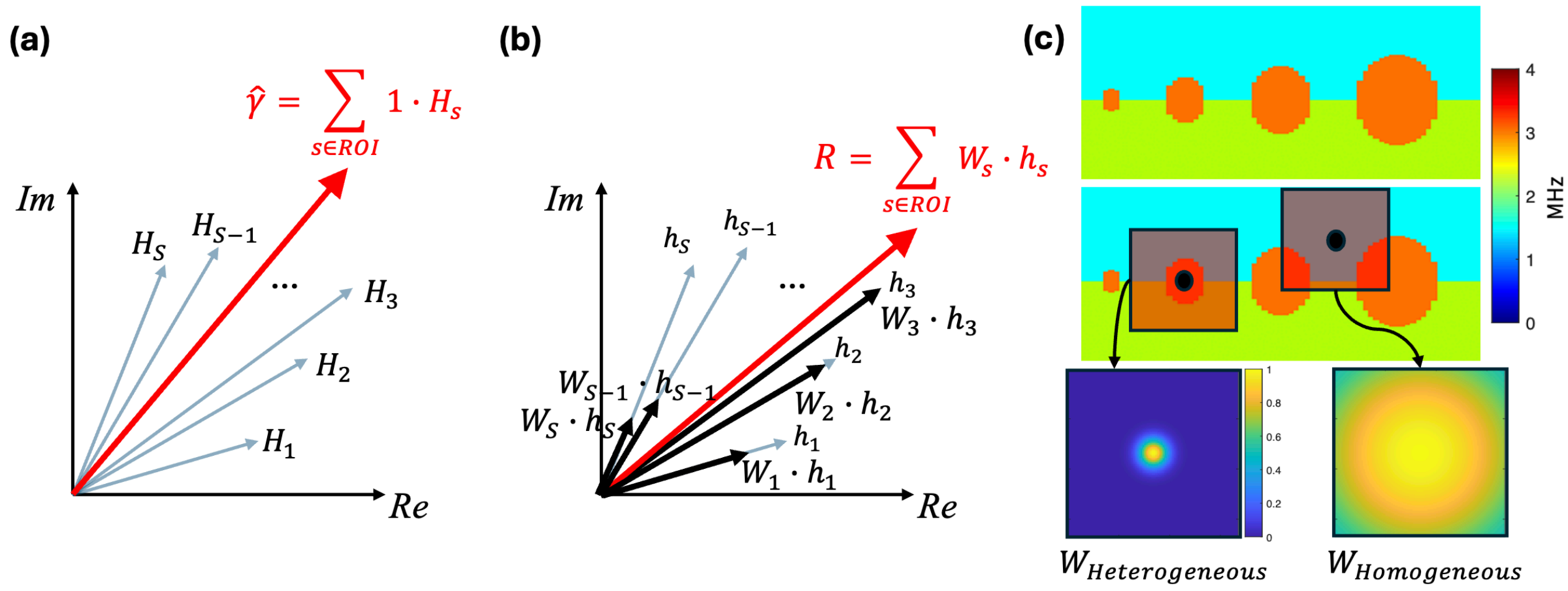

**Figure 1**. Intuitive illustration of calculating autocorrelation. (1) Autocorrelation $\hat{\gamma}$ can be interpreted as vector summation of $H(m,n)|_{m',n'}$ (denoted as $H_h$) in real and imaginary space. All $H_h$ vectors within the ROI contributes equally when calculating $\hat{\gamma}$. (2) Our proposed adaptive autocorrelation $R$, obtained by weighted summation. (c) Illustration of example weight vectors for heterogeneous and homogeneous area.

From the two-dimensional autocorrelator (eq 3), we can estimate mean Doppler frequency $\langle F \rangle$ with $m' = 0$ and $n' = 1$ and mean RF frequency $\langle f \rangle$ with $m' = 1$ and $n' = 0$ [15]:

$$\langle F \rangle \cong \frac{1}{2\pi} \tan^{-1} \left\{ \frac{Im[\hat{\gamma}(0,1)]}{Re[\hat{\gamma}(0,1)]} \right\} \quad \text{(eq 7)}$$

$$\langle f \rangle \cong \frac{1}{2\pi} \tan^{-1} \left\{ \frac{Im[\hat{\gamma}(1,0)]}{Re[\hat{\gamma}(1,0)]} \right\} \quad \text{(eq 8)}$$

The eq 7 and eq 8 can be derived into:

$$\langle f \rangle \cong \frac{1}{2\pi} \tan^{-1} \left\{ \frac{\sum_{m=0}^{M-2} \sum_{n=0}^{N-1} [Q(m,n)I(m+1,n) - I(m,n)Q(m+1,n)]}{\sum_{m=0}^{M-2} \sum_{n=0}^{N-1} [I(m,n)I(m+1,n) + Q(m,n)Q(m+1,n)]} \right\}. \quad \text{(eq 9)}$$

$$\langle F \rangle \cong \frac{1}{2\pi} \tan^{-1} \left\{ \frac{\sum_{m=0}^{M-2} \sum_{n=0}^{N-1} [Q(m,n)I(m,n+1) - I(m,n)Q(m,n+1)]}{\sum_{m=0}^{M-2} \sum_{n=0}^{N-1} [I(m,n)I(m,n+1) + Q(m,n)Q(m,n+1)]} \right\} \quad \text{(eq 10)}$$

Here is the blood velocity ($\langle v_{2D} \rangle$) equation from Loupas's method [15] with the mean Doppler frequency $\langle F \rangle$ and mean RF frequency $\langle f \rangle$:

$$\langle v_{2D} \rangle \cong \frac{c}{2} \frac{\frac{1}{2\pi T_S} \langle F \rangle}{2\pi f_{dem} + \frac{1}{2\pi T_S} \langle f \rangle} \quad \text{(eq 11)}$$

This method enables more precise measurement of blood flow velocity by (1) incorporating mean RF frequency $\langle f \rangle$ and (2) employing a 2D range gate that includes both mean Doppler frequency $\langle F \rangle$ and mean RF frequency $\langle f \rangle$, compared to the traditional 1D Doppler approach which utilizes only $\langle f \rangle$ and 1D range gate.

## 2. 3. Adaptive Frequency Estimator

For tissue characterization to detect pathological tissue changes or differentiate between distinct tissue types, we propose a new RF frequency estimator by refining the autocorrelation equation to make the eq 3 and eq 6 as weighted summation to reflect spatial tissue distribution, meaning that tissues characterized as relatively similar signatures with the tissues at the center of the 2D range gate have higher weights than relatively dissimilar tissues. The term "2D range gate" refers to a region containing samples in both axial (temporal) and lateral directions. In contrast, the 2D range gate for blood flow velocity refers to regions containing axial (fast time) and ensemble (slow time) directions. We leverage the concepts of frequency estimation from Loupas's blood velocity measurement [15], where the concepts include (1) using autocorrelation to estimate mean RF frequency and (2) employing a 2D range gate to achieve more accurate frequency measurement. **Table 1** compared the frequency estimation methodologies of Kasai's color Doppler imaging, Loupas's 2D autocorrelation, and our proposed axial frequency estimator. Loupas's approach measures frequency in both the axial and ensemble directions with the 2D range gate. However, we employ the 2D range gate spanning the axial and lateral directions, but measure only the RF frequency in the axial direction. The purpose of employing the 2D range gate here is to utilize spatial frequency distribution for more accurate estimation, since tissue clusters typically exhibit a circularly spread distribution in a 2D space.

| | **Color Doppler by Kasai** | **2D autocorrelation by Loupas** | **Proposed axial frequency estimator** |
|---|---|---|---|
| **Goal** | Blood flow imaging | Blood flow imaging | Tissue characterization |
| **Frequency estimation** | Frequency shift from blood flow | Frequency shift from blood flow | Axial frequency shifted by pathological or tissue changes |
| **Range gate, window** | 1D range gate (ensemble) | 2D range gate (axial, ensemble) | 2D range gate (axial, lateral) |
| **autocorrelation** | lag 1 at ensemble | Lag 1 at axial, lag 1 at ensemble | Lag 1 at axial |
| **Weight for autocorrelation** | Equal contribution for all samples in range gate | Equal contribution for all samples in range gate | Higher weight for similar tissues with tissue at ROI center; lower weight for dissimilar tissues |

**Table 1**. Comparison of the frequency estimated: Color Doppler [16], 2D autocorrelation [15], and proposed axial frequency estimator.

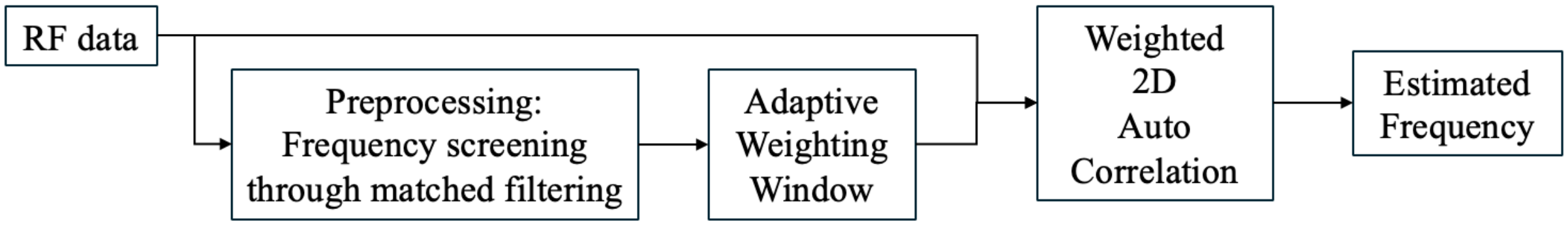

**Figure 2.** Illustration of our proposed axial frequency estimator.

The axial RF frequency $\langle f \rangle$ can be given by:

$$\langle f \rangle \cong \frac{1}{2\pi} \tan^{-1} \left\{ \frac{Im[R(m',l')]}{Re[R(m',l')]} \right\} = \frac{1}{2\pi} \tan^{-1} \left\{ \frac{Im[R(1,0)]}{Re[R(1,0)]} \right\} \quad \text{(eq 12)}$$

where $m'$ and $l'$ represent lags for autocorrelation $R(m', l')$ in the discrete axial (time) and lateral direction, respectively. For our axial frequency estimator, we use $R(1,0)$ with $m' = 1$ and $l' = 0$. We defined $h(m, l)$ as:

$$h(m, l) \equiv IQ(m, l) \cdot IQ^*(m + 1, l) \quad \text{(eq 13)}$$

where IQ represents the in-phase and quadrature data combined as complex numbers. Here, for each $(m, l)$, we updated the autocorrelation equation at lag 1 in axial direction to include weighted summation:

$$R(1, 0) = \sum_{m=0}^{M-2} \sum_{l=0}^{L-1} W(m, l) \cdot [IQ(m, l) \cdot IQ^*(m + 1, l)]$$

$$= \sum_{m=0}^{M-2} \sum_{l=0}^{L-1} W(m, l) \cdot h(m, l) \quad \text{(eq14)}$$

where $W(m, l)$ represents the weight for each $(m, l)$. In the traditional autocorrelation (eq 6), $W(m, l)$ is set to 1 for all samples in a range gate. In our approach, to enhance the precision of our frequency measurements, we adaptively sum $h(m, l)$ based on the frequency distribution within a box ROI (2D range gate), as shown in **Figure 1 (b)**. Thus, our approach requires steps to determine $W(m, l)$ based on spatial tissue distribution. We first screen the frequency distribution using matched filtering, then determine $W(m, l)$, and finally perform a weighted summation for autocorrelation, as shown in **Figure 2**. $W(m, l)$ is defined as a 2D circularly symmetric Gaussian distribution with a standard deviation $\sigma(m, l)$. Since the shape of a tissue or tissue cluster tends to resemble circular forms rather than rectangular boxes, we model the shapes as a 2D circularly symmetric Gaussian distribution and control the sizes by adjusting the $\sigma(m, l)$. We assume that homogeneous tissues within a 2D range gate yield a lower variance in frequency distribution, while heterogeneous tissues result in a greater variance. To quantify homogeneity to determine $W(m,l)$ with $\sigma(m,l)$ based on frequency distribution, we screened frequency through H-scan matched filter analysis [14], resulting in the frequency ($\hat{f}_{Matched}$) for entire images. For all $\{(m,l)\}$ in a 2D range gate, we averaged the frequency to obtain average frequency ($\bar{f}_{Matched}(m,l)$) within a 2D range gate ($ROI_{(m,l)}$) centered at the sample coordinate of $(m,l)$:

$$\bar{f}_{Matched}(m,l) = \frac{1}{S}\sum_{(i,j)\in ROI_{(m,l)}} \hat{f}_{Matched}(i,j) \qquad \text{(eq 15)}$$

where S is the number of samples within a 2D $ROI_{(m,l)}$. We quantified the heterogeneity within $ROI_{(m,l)}$ as the averaged absolute difference ($\Delta$) between $\bar{f}_{Matched}(m,l)$ and $\hat{f}_{Matched}(i,j)$, where $(i,j) \in \mathrm{ROI}_{(m,l)}$, given by:

$$\Delta(m,l) = \frac{1}{S}\sum_{(i,j)\in ROI_{(m,l)}} \left|\bar{f}_{Matched}(m,l) - \hat{f}_{Matched}(i,j)\right|. \qquad \text{(eq 16)}$$

The $\Delta(m,l)$ representing heterogeneity is linearly mapped to $\sigma(m,l)$. For regions of lower heterogeneity or higher homogeneity, we assigned higher $\sigma(m,l)$ for $W(m,l)$, while areas of greater heterogeneity are assigned for lower $\sigma(m,l)$. **Figure 1 (c)** shows examples of 2D maps of $W(m,l)$ for representative heterogeneous and homogeneous areas. By using $W(m,l)$ and combining equations (eq 12) and (eq 14), we obtain RF frequency $\langle f \rangle$ at $(m,l)$:

$$\langle f \rangle|_{(m,l)} \cong \frac{1}{2\pi}\tan^{-1}\left\{\frac{\sum_{(i,j)\in ROI_{(m,l)}} Im[W(i,j)\cdot h(i,j)]}{\sum_{(i,j)\in ROI_{(m,l)}} Re[W(i,j)\cdot h(i,j)]}\right\}. \qquad \text{(eq 17)}$$

# 3. MATERIALS AND METHODS

## 3. 1. Simulation

To evaluate the accuracy of the frequency estimators using a phantom with reference known frequencies, we generated a four-layered medium with circular-shaped inclusions, where each homogeneous medium and inclusion has specific frequencies as shown in **Figure 3**. To generate the validation phantom, we simulated a homogeneous medium using Field II [17] with 0 dB/MHz/cm attenuation to avoid frequency shift due to attenuation. The simulation parameters are summarized in **Table 2**. Using the RF data acquired from Field II simulation, we artificially shifted frequencies to define the layered medium and inclusions by adapting the down-mixing method from IQ demodulation, given as:

$$IQ(t) = RF\ (t) \cdot \exp(-\hat{j}2\pi f_{shift}t). \tag{18}$$

where $f_{shift}$ is an artificial frequency shift from transmission frequency of 3.15 MHz.

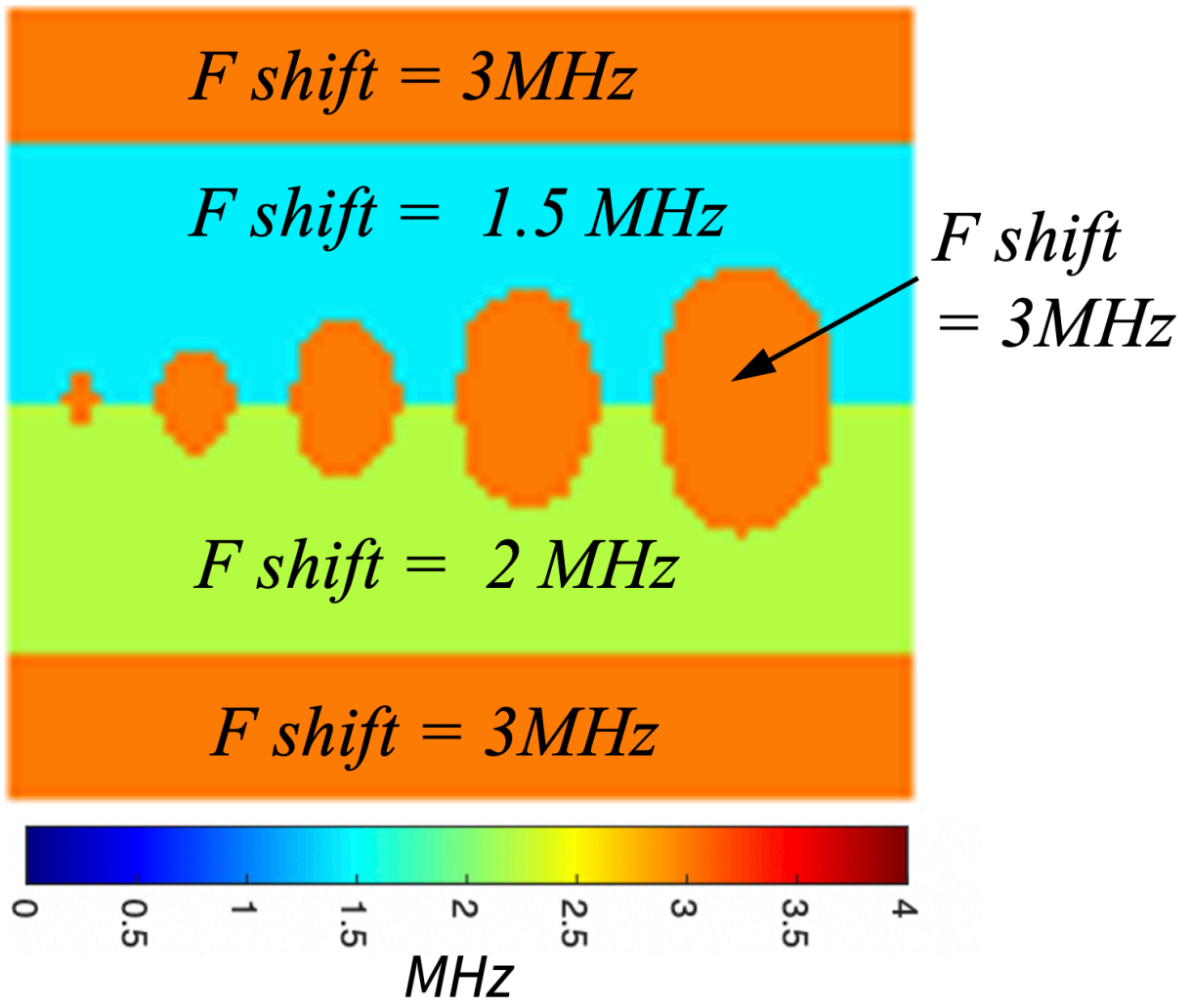

**Figure 3.** Validation phantom simulated by Field II and artificial frequency shift. Transmission frequency = 3.15 MHz.

**Table 2**. Field II simulation properties

| Property | Value |
|---|---|
| Transmit Frequency | 3.15 MHz |
| Attenuation Coefficient | 0 dB/MHz/cm |
| Cycles | 2 |
| Sampling Frequency | 100 MHz |
| # Elements | 129 |
| Kerf | 24 $\mu$m |
| Pitch | 513 $\mu$m |

### 3. 2. In vivo Study

To assess the performance of the adaptive frequency estimator in the quantitative ultrasound field, we applied it to human liver subjects, which include normal livers and hepatic steatosis cases. The *in vivo* study protocol was approved by the Stanford University School of Medicine Review Board. Patients were enrolled and underwent liver MRI scans at the Stanford Hospital using 3.0 T scanners (Discovery MR750, GE Medical System, Waukesha, WI). The human subjects include those with normal livers and those with MASLD. Magnetic resonance imaging-estimated proton density fat fraction (PDFF) was obtained to stage hepatic steatosis: Normal (S0) < 5% < S1 < 10% < S2 < 20% < S3. A Philips EPIQ7 US machine equipped with a C5-1 transducer (Philips Healthcare, Bothell, WA) was used to scan the livers within 14 days of the MRI study. The mean time gap between the MRI and ultrasound examinations was 3.93 days with a standard deviation of 5.25 days, and 55% of the scans were conducted on the same day.

### 3. 3. Validation

To evaluate the performance of our proposed adaptive frequency estimator, we compared it with currently available frequency estimators, such as STFT, the matched filter based estimator from H-scan [14], and 2D autocorrelation without adaptive weighting. Additionally, we investigated median-filtered STFT and median-filtered matched filter. The same 2D kernel size was applied to the approaches, which require a window for measurements, such as median filtered STFT/matched filter, autocorrelation, and the proposed method. The frequency estimators were applied to the simulated phantom with known reference frequencies, results of which were evaluated by calculating root mean squared error (RMSE) with the reference frequencies. Moreover, to assess the potential use of our frequency estimator for quantitative ultrasound, we applied our frequency estimator to human steatosis subjects and evaluated improvement from the currently available H-scan to our approach.

# 4. RESULTS

## 4. 1. Simulation

We evaluated the frequency estimators using the simulated phantom with four-layer medium and circular inclusions (**Figure 3**). **Figure 4 (a)** shows the ground truth map with known frequency distribution. We defined four ROIs, denoted by the boxes in **Figure 4 (a)**, to compare the frequency estimators. ROI 1, 2, 3, and 4 represent the entire image, relatively homogeneous medium including boundaries, an area requiring delineation of small targets, and homogeneous medium, respectively. In **Figure 4 (b) and (c)**, we compared RMSE measures and imaging results, respectively, for the following frequency estimators: STFT, matched filter, median filtered STFT, median filtered matched filter, 2D autocorrelation, and the proposed estimator. As shown in **Figure 4 (b)**, the proposed adaptive frequency estimator resulted in the lowest RMSE for all ROIs, including homogeneous and heterogeneous areas. **Figure 4 (c)** displays magnified images for the 4 ROIs obtained by the frequency estimators. The STFT and matched filter showed sharper boundary area transitions but a salt-and-pepper noise texture. The median-filtered STFT, median-filtered matched filter, and the 2D autocorrelator results showed reduced noise at the expense of boundary sharpness. Especially, for ROI 3, the smallest target on the left disappeared in measuring frequencies with median-filtered STFT and median-filtered matched filter. The proposed approach shows low estimator variance in the homogeneous area and sharper boundary delineation.

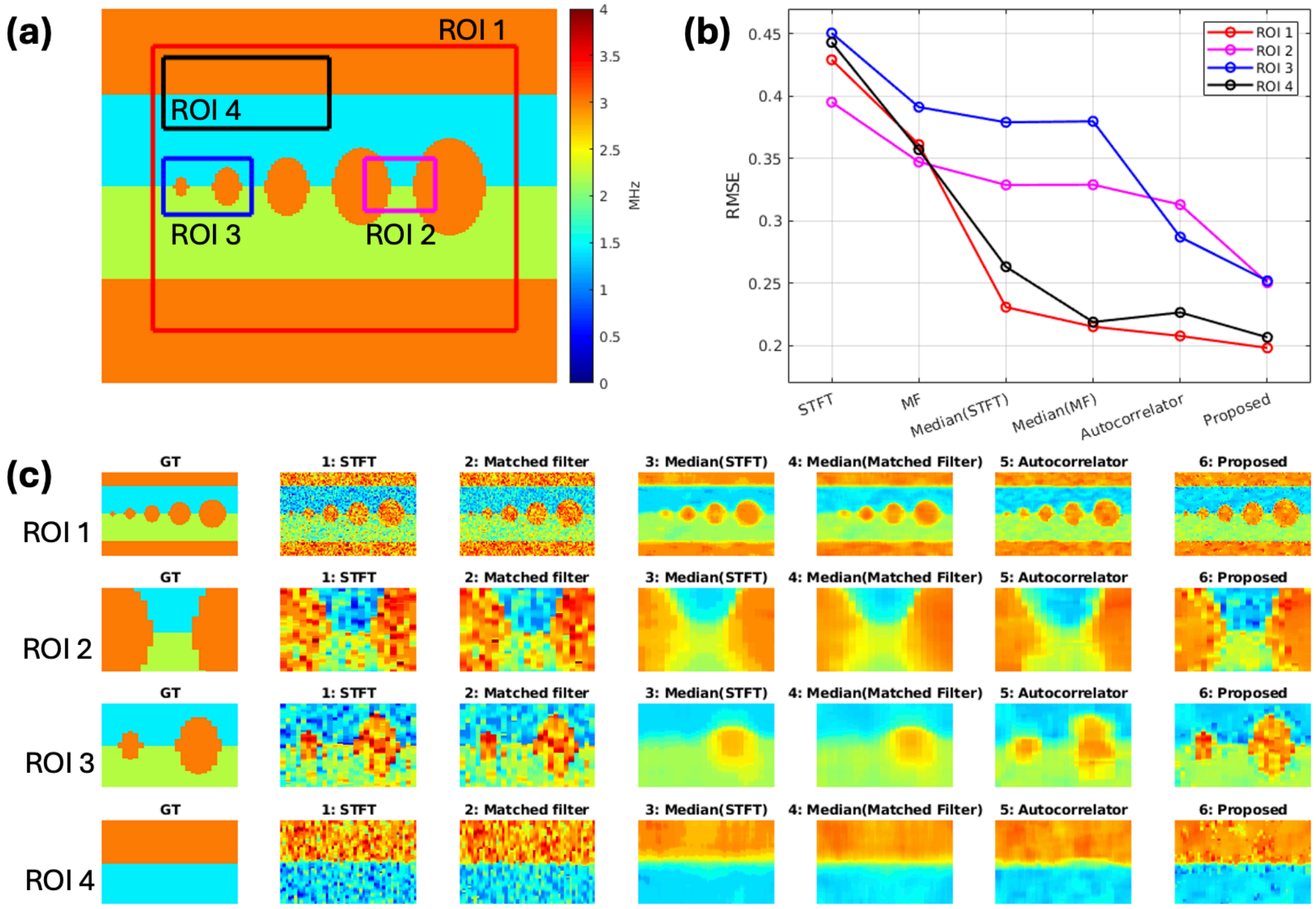

**Figure. 4.** Simulated phantom results obtained by frequency estimators. (a) Ground truth: Simulated phantom with known frequency distribution. Four ROIs are denoted by the boxes. (b) Root mean squared error (RMSE) for the four ROIs. (c) Magnified frequency images for the four ROIs. *GT: ground truth, STFT: short-time Fourier transform, Median (STFT): median filtered STFT, Median (MF): median filtered matched filter.

### 4. 2. H-scan *in vivo* study

**Figure 5** shows B-mode and H-scan images using matched filter [14] with the H-scan binarized colormap, matched filter [14] with jet colormap, and our adaptive frequency estimator with jet colormap, from left to right. **Figure 5 (a) and (b)** show representative cases of normal and stage 3 steatotic human liver, respectively. The left boxes on the B-mode images of **Figure 5 (a) and (b)** are compared in **Figure 5 (c)**; left and right cases in **Figure 5 (c)** represent normal and steatotic livers, respectively. The right box area in the B-mode image of **Figure 5 (b)** is magnified and presented in **Figure 5 (d)**, which illustrates the contrast between steatosis tissues, skin layer, and the gall bladder. Also, **Figure 5** shows the color bars: the almost-binarized red-blue H-scan colormap and a jet colormap, both indicating larger versus smaller scatterers, and a rightmost jet colormap indicating the measured frequency. When comparing normal and steatotic tissues, H-scan images exhibited greater blue-red contrast than the brightness contrast in B-mode, as highlighted in **Figure 5 (c)**. The binarized colormap optimized for H-scan enabled the display of relatively more blue and red pixels for normal and steatotic liver, respectively. However, the measures still contain the salt-and-pepper noise texture, and thus applying the jet colormap to the H-scan matched-filter estimator displays noisy frequency measures (**Figure 5** third column), which hinders clear visual differentiation between normal and steatotic liver. The non-binarized jet map allows comparison of the magnitude of the frequency shift with better resolution than the binarized H-scan colormap, but the noisy texture pattern makes the visual comparison between pathologically different cases difficult. As shown in **Figure 5** fourth column, our proposed adaptive frequency estimator reduces the noise pattern while avoiding loss in resolution of the H-scan image, without requiring a binarized colormap for H-scan. Thus, it shows better visual differentiation between tissues, such as differentiation (1) between normal and steatotic liver (**Figure 5 (c)**) and (2) of steatotic liver from gallbladder/skin layer (**Figure 5 (d)**), while simultaneously providing a quantitative value for tissue characterization. Moreover, H-scan is typically interpreted only quantitatively as relatively larger and smaller scatterers, which is ambiguous and lacks a quantitative value. Our results show a clear quantitative value of frequency measures.

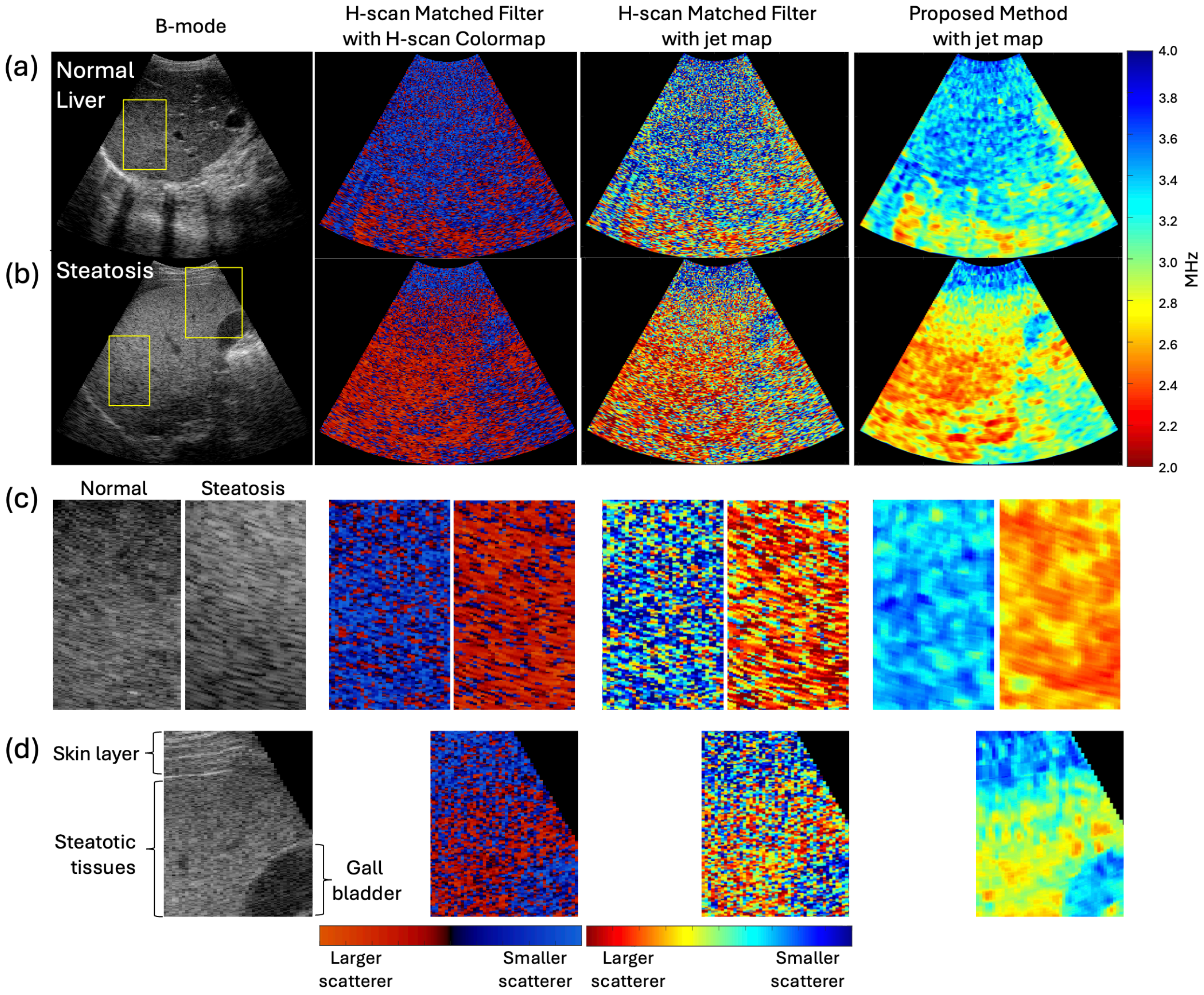

**Figure 5.** *In vivo* frequency estimation results applied to H-scan matched filter analysis and our proposed method for a steatotic liver. (a) Normal liver. (b) Stage 3 steatotic liver. (c) Comparison of H-scan images to differentiate between normal and steatosis. (d) Comparison of H-scan images to differentiate distinct tissues, such as skin layer, steatotic tissues, and the gall bladder.

# 5. DISCUSSION

In this study, we developed an adaptive frequency estimator for H-scan tissue characterization, which can improve the current H-scan thereby providing less noisy measures and better illustration of pathological conditions of tissues. Specifically, the proposed approach enabled smooth textures in areas of homogeneous tissue and high contrast in boundaries between different tissue types. The frequency estimation accuracy was evaluated using a simulated phantom with known frequency distribution, which demonstrated that the proposed estimator resulted in the lowest RMSE with the reference phantom compared to currently available frequency estimators. The proposed frequency estimator was applied to H-scan imaging for *in vivo* human livers, resulting in improved tissue characterization in terms of less noisy liver texture, better differentiation between normal and steatotic cases, and clearer boundary delineation between different tissue types, such as hepatic tissues, skin, and gall bladder, compared to currently available H-scan [14].

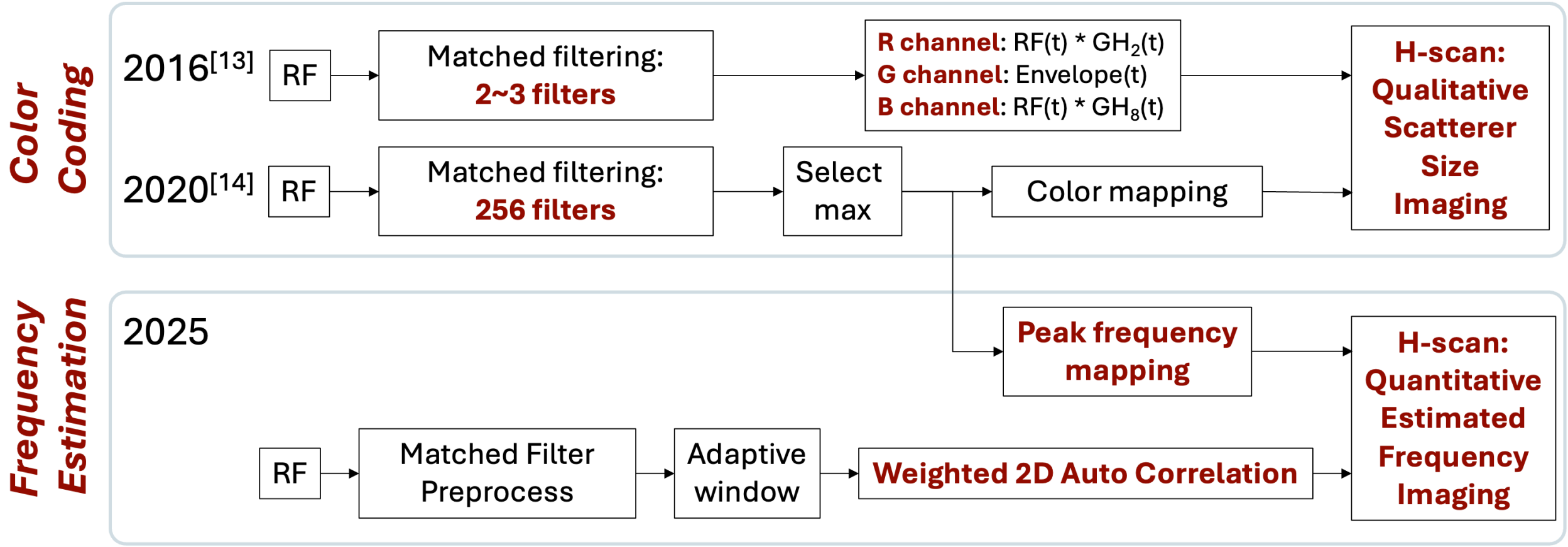

**Figure 6.** H-scan history

H-scan history is summarized in **Figure 6**. H-scan was first proposed [13] as a matched filter analysis with a simple implementation using two Gaussian weighted Hermite functions, of which convolution with RF data detects relatively smaller and larger scatterers. The convolutions were assigned to the red and blue channels, and envelope data was assigned to the green channel. The resulting RGB image was termed H-scan. This originally proposed H-scan was applied to thyroid lesions [18]. The measures from normal, benign, and malignant showed a statistically significant difference, but H-scan images for benign and malignant were barely differentiated visually due to the noisy pattern in color imaging. To improve the H-scan color resolution, a new approach of H-scan analysis [14] was proposed, which utilized 256 matched filters. Since the RGB approach is not capable of utilizing 256 convolution results, it proposed to use mapping from the 256 matched filtering results to a 256-level color-scale with the proposed H-scan color bar, shown in **Figure 5**. The H-scan colormap distributes from red to blue, and the red-blue colormap was optimized to be almost binarized to visually suppress the noisy pattern. This new H-scan approach with the optimized H-scan colormap is capable of providing improved visual differentiation of different pathological conditions, such as liver fibrosis [4], liver steatosis [2], pancreatic cancer metastasis [6], melanoma [9], breast cancer [12], and kidney fibrosis [19], etc. However, both approaches provide qualitative information on relative sizes of scatterers, meaning more red and blue pixels represent smaller and larger scatterers, respectively. In this study, we aimed to provide quantitative measures of H-scan while reducing the noisy measures. We updated H-scan approach to visualize frequency estimates instead of relative sizes, and moreover for more accurate measures than currently available frequency estimators, we developed a new frequency estimator, which is appropriate for tissue characterization. We proposed the refined autocorrelation with weighted summation, which adaptively update a 2D weight map for the 2D range gate in autocorrelation based on 2D spatial tissue distribution. Therefore, our adaptive estimator enabled a smooth texture pattern for homogeneous tissue areas but clear delineation for small targets or tissue boundaries, while currently available estimators have limitations of trade-off between homogeneous texture pattern and clear boundary delineation.

The original H-scan approaches provide measurements as percentages of blue pixels or color index among 256 levels, where higher blue percentages or higher color levels indicate relatively larger scatterer sizes. However, these relative scales are ambiguous because they cannot directly measure scatterer sizes. Here, our H-scan framework applied to images obtained by the Philips EPIQ7 US system with a C5-1 transducer (Philips Healthcare, Bothell, WA) provides quantitative frequency measurement; the normal liver exhibited a frequency of 3.6 MHz, while the steatotic liver showed a 2.4 MHz. These measures can be used to screen new patients with the same US system. For instance, patients exhibiting H-scan frequency measurements approximately 3.6 MHz are more likely to present with a normal liver condition, while measures shifted to 2.4 MHz suggest a higher likelihood of steatosis. However, the baseline frequencies for normal and steatosis (3.6 MHz and 2.4 MHz in this study) depend on the US system's transmission frequency and bandwidth. Thus, the baseline needs to be initialized for each system and transducer.

To validate our H-scan utilizing the adaptive frequency estimator, we applied it to hepatic steatosis, demonstrating improved visual differentiation between normal and steatotic livers compared to the H-scan [14], as shown in **Figure 5**. Steatosis generally develops homogeneous infiltration of fat inclusions across wide areas of liver tissue, rather than localized tissue changes such as liver lesions. Therefore, our H-scan imaging showed suppressed salt-and-pepper noise texture in this steatosis study. Furthermore, we anticipate that our approach could enable high contrast for tissue boundary delineation, including small inclusions, when applied to detect benign or malignant liver lesions.

## 6. CONCLUSIONS

We have developed an adaptive frequency estimator for H-scan tissue characterization, thereby (1) quantifying H-scan measures, (2) enhancing frequency estimation accuracy by refining the 2D autocorrelation to be more suitable for tissue characterization, and (3) improving H-scan imaging results. In our frequency estimation framework, the matched filter first screens the spatial distribution of frequencies, and then the refined 2D autocorrelation estimates the local frequencies with adaptively determined window weights based on 2D spatial frequency distribution reflecting tissue properties. The H-scan utilizing our frequency estimator suppressed noisy texture in homogeneous regions, preserved contrast at the boundaries of different tissue types, and maintained clarity in the delineation of small textures. Our approach was applied to *in vivo* human hepatic steatosis subjects, demonstrating a suppressed noise pattern, improved visual differentiation between normal and steatotic livers, improved delineation of tissue boundaries (e.g., liver tissues from gallbladder/skin layer). Therefore, we expect that the enhanced H-scan imaging with the proposed frequency estimator will assist clinicians in reading and interpreting H-scan ultrasound images more easily for diagnosis.

## Acknowledgments

This work was supported by National Institutes of Health grant R01-EB027100 and Stanford School of Medicine Dean's Fellowship.

[1] A. M. Pirmoazen, A. Khurana, A. M. Loening *et al.*, "Diagnostic performance of 9 quantitative ultrasound parameters for detection and classification of hepatic steatosis in nonalcoholic fatty liver disease," *Investigative Radiology,* vol. 57, no. 1, pp. 23-32, 2022.

[2] J. Baek, S. S. Poul, L. Basavarajappa *et al.*, “Clusters of ultrasound scattering parameters for the classification of steatotic and normal livers,” *Ultrasound in Medicine & Biology,* vol. 47, no. 10, pp. 3014-3027, 2021.

[3] J. Baek, L. Basavarajappa, K. Hoyt *et al.*, “Disease-specific imaging utilizing support vector machine classification of H-scan parameters: assessment of steatosis in a rat model,” *IEEE Transactions on Ultrasonics, Ferroelectrics, and Frequency Control,* vol. 69, no. 2, pp. 720-731, 2021.

[4] J. Baek, S. S. Poul, T. A. Swanson *et al.*, “Scattering Signatures of Normal Versus Abnormal Livers with Support Vector Machine Classification,” *Ultrasound in Medicine and Biology,* vol. 46, no. 12, pp. 3379-3392, Dec, 2020.

[5] J. Baek, T. A. Swanson, T. Tuthill *et al.*, “Support vector machine (SVM) based liver classification: fibrosis, steatosis, and inflammation,” *Proceedings of the 2020 Ieee International Ultrasonics Symposium (Ius)*, 2020.

[6] J. Baek, R. Ahmed, J. Ye *et al.*, “H-Scan, Shear Wave and Bioluminescent Assessment of the Progression of Pancreatic Cancer Metastases in the Liver,” *Ultrasound in Medicine and Biology,* vol. 46, no. 12, pp. 3369-3378, Dec, 2020.

[7] J. Baek, E. Hysi, X. He *et al.*, "Detecting kidney fibrosis using H-scan." pp. 1-3.

[8] J. Baek, S. S. Qin, P. A. Prieto *et al.*, "H-scan imaging and quantitative measurement to distinguish melanoma metastasis." pp. 1-4.

[9] J. Baek, S. S. Qin, P. A. Prieto *et al.*, “H-Scan Discrimination for Tumor Microenvironmental Heterogeneity in Melanoma,” *Ultrasound in Medicine & Biology,* vol. 50, no. 2, pp. 268-276, 2024.

[10] J. Baek, A. El Kaffas, A. Kamaya *et al.*, “Multiparametric quantification and visualization of liver fat using ultrasound,” *WFUMB Ultrasound Open,* vol. 2, no. 1, pp. 100045, 2024.

[11] J. Baek, S. Sanabria, I. Oyarzabal *et al.*, "A multi-parametric model for progression of metabolic dysfunction-associated steatohepatitis (MASH) in humans." pp. 1-3.

[12] J. Baek, A. M. O’Connell, and K. J. Parker, “Improving breast cancer diagnosis by incorporating raw ultrasound parameters into machine learning,” *Machine Learning: Science and Technology,* vol. 3, no. 4, pp. 045013, 2022.

[13] K. Parker, “Scattering and reflection identification in H-scan images,” *Physics in Medicine & Biology,* vol. 61, no. 12, pp. L20, 2016.

[14] K. J. Parker, and J. Baek, “Fine-tuning the H-scan for discriminating changes in tissue scatterers,” *Biomedical Physics & Engineering Express,* vol. 6, no. 4, Jul, 2020.

[15] T. Loupas, J. Powers, and R. W. Gill, “An axial velocity estimator for ultrasound blood flow imaging, based on a full evaluation of the Doppler equation by means of a two-dimensional autocorrelation approach,” *IEEE transactions on ultrasonics, ferroelectrics, and frequency control,* vol. 42, no. 4, pp. 672-688, 1995.

[16] C. Kasai, K. Namekawa, A. Koyano *et al.*, “Real-time two-dimensional blood flow imaging using an autocorrelation technique,” *IEEE Transactions on sonics and ultrasonics,* vol. 32, no. 3, pp. 458-464, 1985.

[17] J. A. Jensen, "Field: A program for simulating ultrasound systems."

[18] R. G. Gary, R. Laimes, J. Pinto *et al.*, “H-scan analysis of thyroid lesions,” *Journal of Medical Imaging,* vol. 5, no. 1, pp. 013505, 2018.

[19] E. Hysi, J. Baek, A. Koven *et al.*, “A first-in-human study of quantitative ultrasound to assess transplant kidney fibrosis,” *Nature Medicine*, pp. 1-9, 2025.